\documentclass[12pt, onecolumn]{IEEEtran}
\usepackage{amsmath, amsthm, amssymb}
\usepackage{commands11}
\usepackage{cite}
\usepackage[margin=.85in]{geometry}

\newtheorem{theorem}{Theorem}
\newtheorem{lemma}{Lemma}

\title{Information and Estimation over Binomial and Negative Binomial Models}
\author{Dongning Guo\\
Department of Electrical Engineering and Computer Science\\
Northwestern University\\
Evanston, IL 60208, USA}

\begin{document}
\maketitle

\begin{abstract}
  In recent years, a number of results have been developed which
  connect information measures and estimation measures under various
  models, including, predominently, Gaussian and Poisson models.  More
  recent results due to Taborda and P\'erez-Cruz relate the relative
  entropy to certain mismatched estimation errors in the context of
  binomial and negative binomial models, where, unlike in the case of
  Gaussian and Poisson models, the conditional mean estimates concern
  models of different parameters than those of the original model.  In
  this note, a different set of results in simple forms are
  developed for binomial and negative binomial models, where the
  conditional mean estimates are produced through the original models.
  The new results are more consistent with existing results for
  Gaussian and Poisson models. 
\end{abstract}

\section{Introduction}

Since a simple differential relationship between the mutual
information and the minimum mean square error over a scalar Gaussian
model was discovered~\cite{GuoSha05IT}, a number of similar results
have been developed for several other models, e.g., Poisson
models~\cite{GuoSha08IT, AtaWei12IT}.  
In the context of Gaussian and Poisson models, it has also been found that the
relative entropy can be expressed as the integral of the increase of
the estimation error due to mismatched prior
distribution~\cite{Verdu10IT, Weissm10IT, AtaWei12IT}. 

More recently, Taborda and P\'erez-Cruz~\cite{TabPer12ISIT} 
developed results in the 
context of binomial and negative binomial models.  The key result
expresses the derivative of the relative entropy of two output
distributions of the same binomial (or negative binomial) model
induced by two different inputs in terms of certain mismatched estimation
errors.  As in the case of Gaussian and Poisson models, the errors
concern conditional mean estimates of the input given the output.
However, the conditional mean is produced using binomial (or negative
binomial) models with modified parameters. In particular, in case of
the binomial model, the modified model is of one fewer trial than the
original model; in case of the negative model, the modified model is
of one more failure than the original model.

In this note, we develop a different set of results concerning
essentially the same binomial and negative binomial models as
in~\cite{TabPer12ISIT}. 
The results put the derivative of the relative entropy in a simple
form concerning some average difference of conditional mean estimates
due to mismatched prior distribution.  In contrast to
results of~\cite{TabPer12ISIT}, the conditional mean estimates here
are based on the original binomial and negative binomial models.
The results are thus more consistent with existing results for
Gaussian and Poisson models.

\clearpage
\section{The Binomial Model}

The binomial model is based on the binomial distribution, Binomial($n,q$), which describes the probability of having $k$ successful trials in $n$ independent Bernoulli trials, each with probability $q$ to succeed:
\begin{align}
  P( Y=k ) = \binom{n}{k} q^k (1-q)^{n-k}, \quad k=0,\dots,n.
\end{align}
With some hindsight, we let the binomial model be a random
transformation from a random variable $X$ which takes its value on
$(a,\infty)$ to another random variable $Y$, where, conditioned on
$X=x$, $Y$ follows the distribution of Binomial($n,a/x$).  The
conditional probability mass function is given by 
\begin{align} \label{eq:PYX}
  p_{Y|X}(y|x) = \binom{n}{y} \left(\frac{a}x\right)^y \left(1-\frac{a}x\right)^{n-y},
  \quad y=0,\dots,n.
\end{align}
The variables $X$ and $Y$ are viewed as the input and output of the
binomial model.  Here the input $X$ controls the probability of
success of an individual Bernoulli trial, namely, for fixed $X$, the
success-to-failure ratio is $a:X$.  The larger $X$ is, the fewer
trials succeed on average.  The parameter $a$ is viewed as a scaling
of the input. 

If the prior distribution of the input $X$ is $P_X$, the corresponding
output distribution is denoted by $P_Y$; if the prior distribution of
$X$ is $Q_X$, the corresponding output distribution is denoted by
$Q_Y$.  Throughout this note, we use $\expect{\cdot}$ and $\expcnd{\cdot}{\cdot}$ to denote expectation and conditional expectation under distribution $P$, whereas we use
$\expsub{Q}{\cdot}$ and $\expsubcnd{Q}{\cdot}{\cdot}$ to denote expectation and conditional expectation under distribution $Q$.  Thus the conditional mean of $X$ given $Y$ is denoted by $\expcnd{X}{Y}$ under distribution $P$ and by $\expsubcnd{Q}{X}{Y}$ under distribution $Q$.

We also define the following function
\begin{align} \label{eq:guo(t)}
  g(t) = t - 1 - \log t.
\end{align}
This function is convex on $(0,\infty)$, and achieves its unique
minimum, 0, at $t=1$.
For two positive numbers, $x$ and $\hat{x}$, the function
$g(x/\hat{x})$ can be viewed as a measure of their difference, in the
sense that it is always nonnegative, and that it is equal to 0 if and
only if $x=\hat{x}$. Moreover, $g(x/\hat{x})$ increases monotonically as
$x/\hat{x}$ departs from 1 in either direction of the axis. 

\begin{theorem} \label{th:dadDPQ}
  Let $P_Y$ and $Q_Y$ be the output distribution of the binomial
  model~\eqref{eq:PYX} induced by input distributions $P_X$ and $Q_X$,
  respectively, where $P_X$ and $Q_X$ put no probability mass on
  $(-\infty,a]$.  Then 
  \begin{align} \label{eq:dadDPQ}
    \pd{a} \divergence{P_Y}{Q_Y}
    = \expect{ \frac{Y}a \cdot
      g\left( \frac{\expcnd{X}{Y}-a}{\expsubcnd{Q}{X}{Y}-a} \right) }.
  \end{align}
\end{theorem}

\begin{lemma} \label{lm:dadPY}
  Let $P_Y$ be the probability mass function of the output of the binomial model described by~\eqref{eq:PYX}, where the input follows distribution $P_X$ with zero mass on $(-\infty,a]$.
  For every $y=0,\dots,n$,
  \begin{align} \label{eq:dadPY}
    \pd{a} P_Y(y) = \frac1a \left( y P_Y(y) - (y+1) P_Y(y+1) \right)
  \end{align}
  where we use the convention that $p_Y(n+1)=0$.  The result remains true if $P_Y$ is replaced by $Q_Y$ in~\eqref{eq:dadPY}.
\end{lemma}

\begin{IEEEproof}
  We start with
  \begin{align}\label{eq:PYy}
    P_Y(y) = \expect{
      \binom{n}{y} \left(\frac{a}X\right)^y \left(1-\frac{a}X\right)^{n-y} }.
  \end{align}
  Evidently,
  \begin{align}
    \pd{a} P_Y(y)
    &= \expect{
      \binom{n}{y} \pd{a} \left( 
        \left(\frac{a}X\right)^y \left(1-\frac{a}X\right)^{n-y} \right)} \label{eq:dadPY=} \\
    &= \expect{ \binom{n}{y}
       \left( \pd{a} \left(\frac{a}X\right)^y \right) \left(1-\frac{a}X\right)^{n-y} }
      + \expect{ \binom{n}{y}
        \left(\frac{a}X\right)^y \pd{a} \left(1-\frac{a}X\right)^{n-y} } \\
    &= \frac{y}a \expect{ \binom{n}{y}
      \left(\frac{a}X\right)^y \left(1-\frac{a}X\right)^{n-y} }
      - \frac{n-y}a \expect{ \binom{n}{y}
        \left(\frac{a}X\right)^{y+1} \left(1-\frac{a}X\right)^{n-y-1} } \label{eq:n-y} \\
    &= \frac{y}a P_Y(y) - \frac{y+1}a \expect{ \binom{n}{y+1}
      \left(\frac{a}X\right)^{y+1} \left(1-\frac{a}X\right)^{n-y-1} }. \label{eq:n+1}
  \end{align}
  We note that~\eqref{eq:dadPY=}--\eqref{eq:n-y} hold for $y=0,\dots,n$.  In arriving at~\eqref{eq:n+1}, we use~\eqref{eq:PYy} and the convention that $\binom{n}{n+1}=0$.  In fact the second term in~\eqref{eq:n-y} and the second term in~\eqref{eq:n+1} are both equal to 0 for $y=n$.  Using~\eqref{eq:PYy} again, we arrive at~\eqref{eq:dadPY} from~\eqref{eq:n+1}.

Since~\eqref{eq:dadPY} holds for any input distribution $P_X$, it remains true if $P_X$ is replaced by another distribution $Q_X$, as long as the input is always greater than $a$.
\end{IEEEproof}

Lemma~\ref{lm:dadPY} resembles a result for Gaussian models
in~\cite{GuoSha05IT}, where the derivative with respect to the scaling
parameter translates to the derivative with respect to the output
variable.  For the binomial model, the output is discrete and the
result consists of the difference of the output distribution
(modulated by the variable $y$) in lieu of derivative.

We next prove Theorem~\ref{th:dadDPQ}.

\begin{IEEEproof}[Proof of Theorem~\ref{th:dadDPQ}]
  From
  \begin{align}
    \divergence{P_Y}{Q_Y}
    = \sum^n_{y=0} P_Y(y) \log \frac{P_Y(y)}{Q_Y(y)},
  \end{align}
  it is not difficult to show that
  \begin{align}
    \pd{a}    \divergence{P_Y}{Q_Y}
    &= \sum^n_{y=0} \left( \log\frac{P_Y(y)}{Q_Y(y)} \right) \frac{\diff P_Y(y)}{\diff a}
    - \frac{P_Y(y)}{Q_Y(y)} \frac{\diff Q_Y(y)}{\diff a} \\
    &= a^{-1} (A-B) \label{eq:dadPQAB}
  \end{align}
  where
  \begin{align}
    A
    &= a \sum^n_{y=0} \left( \log\frac{P_Y(y)}{Q_Y(y)} \right) \frac{\diff P_Y(y)}{\diff a} \\
    &= \sum^n_{y=0} \left( \log\frac{P_Y(y)}{Q_Y(y)} \right) (y P_Y(y) - (y+1) P_Y(y+1)) \\
    &= \sum^n_{y=1} \left( \log\frac{P_Y(y)}{Q_Y(y)} \right) y P_Y(y) 
    - \sum^{n-1}_{y=0} \left( \log\frac{P_Y(y)}{Q_Y(y)} \right) (y+1) P_Y(y+1)) \\
    &= \sum^n_{y=1} \left( \log\frac{P_Y(y)}{Q_Y(y)} \right) y P_Y(y) 
    - \sum^n_{y=1} \left( \log\frac{P_Y(y-1)}{Q_Y(y-1)} \right) y P_Y(y) \\
    &= \sum^n_{y=1} y P_Y(y) \log\frac{P_Y(y)Q_Y(y-1)}{P_Y(y-1)Q_Y(y)} 
    \label{eq:A}
  \end{align}
  and
  \begin{align}
    B
    &= a \sum^n_{y=0} \frac{P_Y(y)}{Q_Y(y)} \frac{\diff Q_Y(y)}{\diff a} \\
    &= \sum^n_{y=0} \frac{P_Y(y)}{Q_Y(y)} (y Q_Y(y) - (y+1) Q_Y(y+1)) \\
    &= \sum^n_{y=1} y P_Y(y) 
    - \sum^{n-1}_{y=0} \frac{P_Y(y)}{Q_Y(y)} (y+1) Q_Y(y+1)) \\
    &= \sum^n_{y=1} y P_Y(y) 
    - \sum^n_{y=1} \frac{P_Y(y-1)}{Q_Y(y-1)} y Q_Y(y) \\
    &= \sum^n_{y=1} y P_Y(y) \left( 1- \frac{P_Y(y-1)Q_Y(y)}{P_Y(y)Q_Y(y-1)} \right) .
    \label{eq:B}
  \end{align}
  Moreover, 
  \begin{align}
    P_Y(y-1)
    &= \expect{
      \binom{n}{y-1} \left(\frac{a}X\right)^{y-1} \left(1-\frac{a}X\right)^{n-y+1} } \\
    &= \expect{ \frac{y}{n-y+1}
      \left(\frac{a}X\right)^{-1} \left(1-\frac{a}X\right)
      \binom{n}{y} \left(\frac{a}X\right)^y \left(1-\frac{a}X\right)^{n-y} } \\
    &= \frac{y}{n-y+1} \expcnd{\frac{X}a-1}{Y=y} P_Y(y).
  \end{align}
  Similarly,
  \begin{align}
    Q_Y(y-1)
    &= \frac{y}{n-y+1} \expsubcnd{Q}{\frac{X}a-1}{Y=y} Q_Y(y).
  \end{align}
  Therefore,
  \begin{align} \label{eq:PQ/PQ}
    \frac{P_Y(y-1)Q_Y(y)}{P_Y(y)Q_Y(y-1)}
    &= \frac{\expcnd{X}{Y=y}-a}
    {\expsubcnd{Q}{X}{Y=y}-a}.
  \end{align}
  Plugging~\eqref{eq:PQ/PQ} into~\eqref{eq:A} and~\eqref{eq:B} and subsequently~\eqref{eq:dadPQAB}, we have
  \begin{align}
    \pd{a} \divergence{P_Y}{Q_Y}
    &= a^{-1} \sum^n_{y=1} y P_Y(y) ( T(y) - 1 - \log T(y) )
  \end{align}
  where $T(y)$ is a shorthand for the function defined as the RHS of~\eqref{eq:PQ/PQ}.
  By definition~\eqref{eq:guo(t)}, we have established~\eqref{eq:dadDPQ} in Theorem~\ref{th:dadDPQ}.
\end{IEEEproof}

\clearpage
\section{The Negative Binomial Model}

The negative binomial distribution is defined by the following probability mass function
\begin{align} 
  P(Y=y) = \binom{y+r-1}{y} (1-q)^r q^y, \quad y=0,1,\dots
\end{align}
which is the probability that $y$ successful trials are seen before the $r$-th failure is observed, where the trials are independent Bernoulli trials each with probability $q$ to succeed.  We denote this distribution as $-$Binomial($r,q$).

With some hindsight, we define a negative binomial model based on random transformation from random variable $X$ to random variable $Y$, where, conditioned on $X=x$, $Y$ has distribution $-$Binomial($r,b/(b+x)$).  That is, the random transformation is given by conditional probability mass function
\begin{align} \label{eq:-PYX}
  P_{Y|X}(y|x) = \binom{y+r-1}{y} \left( \frac{x}{b+x} \right)^r 
  \left( \frac{b}{b+x} \right)^y, \quad y=0,1,\dots.
\end{align}

\begin{theorem} \label{th:dbdDPQ}
  Let $P_Y$ and $Q_Y$ be the output distribution of the binomial
  model~\eqref{eq:-PYX} induced by input distributions $P_X$ and $Q_X$,
  respectively, where $P_X$ and $Q_X$ put no probability mass on
  $(-\infty,0]$.  Then 
  \begin{align} \label{eq:dbdDPQ}
    \pd{b} \divergence{P_Y}{Q_Y}
    = \expect{ \frac{Y}b \cdot
      g\left( \frac{\expcnd{X}{Y}+b}{\expsubcnd{Q}{X}{Y}+b} \right) }.
  \end{align}
\end{theorem}
 
\begin{lemma} \label{lm:dbdPY}
  Let $P_Y$ be the probability mass function of the output of the
  negative binomial model described by~\eqref{eq:-PYX}, where the
  input is always positive and follows distribution $P_X$.
  For every $y=0,1,\dots$,
  \begin{align} \label{eq:dbdPY}
    \pd{b} P_Y(y) = \frac1b \left( y P_Y(y) - (y+1) P_Y(y+1) \right).
  \end{align}
  The result remains true if $P_Y$ is replaced by $Q_Y$ in~\eqref{eq:dbdPY}.
\end{lemma}

\begin{IEEEproof}
  We start with
  \begin{align}\label{eq:-PYy}
    P_Y(y) = \expect{
      \binom{y+r-1}{y} \left(\frac{X}{b+X}\right)^r \left(\frac{b}{b+X}\right)^y }.
  \end{align}
  Evidently,
  \begin{align}
    \pd{b} P_Y(y)
    &= \expect{
      \binom{y+r-1}{y} \pd{b} \left( 
        \left(\frac{X}{b+X}\right)^r
        \left(\frac{b}{b+X}\right)^y \right)} \label{eq:-dadPY=} \\
    &= \expect{ \binom{y+r-1}{y} \left[
      \left( \pd{b} \left(\frac{X}{b+X}\right)^r \right)
      \left(\frac{b}{b+X}\right)^y +
        \left(\frac{X}{b+X}\right)^r \pd{b}
        \left(\frac{b}{b+X}\right)^y \right] } \\
    &= \expect{ \binom{y+r-1}{y}
      \left(\frac{X}{b+X}\right)^r \left(\frac{b}{b+X}\right)^y
      \left( \frac{-r}{b+X} + \frac{yX}{b(b+X)} \right) }  \label{eq:yX-r} \\
    &= \expect{ \binom{y+r-1}{y}
      \left(\frac{X}{b+X}\right)^r \left(\frac{b}{b+X}\right)^y
      \left( \frac{y}{b} - \frac{y+r}{b+X} \right) }  \label{eq:y+r} \\
    &= \frac{y}b P_Y(y) - \frac{y+1}b \expect{ \binom{y+r}{y+1}
      \left(\frac{X}{b+X}\right)^r \left(\frac{b}{b+X}\right)^{y+1}
    } \label{eq:y+1} \\
    &= \frac1b \left( y P_Y(y) - (y+1) P_Y(y+1) \right).
  \end{align}
  
  Since~\eqref{eq:dbdPY} holds for any input distribution $P_X$, it
  remains true if $P_X$ is replaced by another distribution $Q_X$, as
  long as the input is always nonnegative.
\end{IEEEproof}

It is interesting to see that~\eqref{eq:dbdPY} is literally identical
to~\eqref{eq:dadPY} if the two parameters $a$ and $b$ are identical.

The proof of Theorem~\ref{th:dbdDPQ} based on Lemma~\ref{lm:dbdPY}
resembles that of Theorem~\ref{th:dadDPQ}.

\begin{IEEEproof}[Proof of Theorem~\ref{th:dbdDPQ}]
  Using similar techniques as in the proof of Theorem~\ref{th:dbdDPQ},
  we arrive at
  \begin{align} \label{eq:dbdDPQT}
    \pd{b} \divergence{P_Y}{Q_Y}
    &= \frac1b \sum^\infty_{y=1} y P_Y(y) ( T(y) - 1 - \log T(y) )
  \end{align}
  where
  \begin{align} \label{eq:T(y)}
    T(y) = \frac{P_Y(y-1)Q_Y(y)}{P_Y(y)Q_Y(y-1)}.
  \end{align}
  Moreover, 
  \begin{align}
    P_Y(y-1)
    &= \expect{
      \binom{y+r-2}{y-1} \left(\frac{X}{b+X}\right)^r
      \left(\frac{b}{b+X}\right)^{y-1} } \\
    &= \expect{ \frac{y}{y+r-1}\cdot\frac{b+X}b
      \binom{y+r-1}{y} \left(\frac{X}{b+X}\right)^r \left(\frac{b}{b+X}\right)^y } \\
    &= \frac{y}{y+r-1} \expcnd{1+\frac{X}b}{Y=y} P_Y(y).
  \end{align}
  Similarly,
  \begin{align}
    Q_Y(y-1)
    &= \frac{y}{y+r-1} \expsubcnd{Q}{1+\frac{X}b}{Y=y} Q_Y(y).
  \end{align}
  Therefore,
  \begin{align} \label{eq:-PQ/PQ}
    T(y)
    &= \frac{\expcnd{X}{Y=y}+b}
    {\expsubcnd{Q}{X}{Y=y}+b}.
  \end{align}
  Theorem~\ref{th:dbdDPQ} is thus established using~\eqref{eq:guo(t)},~\eqref{eq:dbdDPQT} and~\eqref{eq:-PQ/PQ}.
\end{IEEEproof}


\begin{thebibliography}{1}

\bibitem{GuoSha05IT}
D.~Guo, S.~{Shamai (Shitz)}, and S.~Verd\'u, ``Mutual information and minimum
  mean-square error in {G}aussian channels,'' {\em IEEE Trans.\ Inform.\
  Theory}, vol.~51, pp.~1261--1282, Apr. 2005.

\bibitem{GuoSha08IT}
D.~Guo, S.~{Shamai (Shitz)}, and S.~Verd\'u, ``Mutual information and
  conditional mean estimation in {P}oisson channels,'' {\em IEEE Trans.\
  Inform.\ Theory}, vol.~54, pp.~1837--1849, May 2008.

\bibitem{AtaWei12IT}
R.~Atar and T.~Weissman, ``Mutual information, relative entropy, and estimation
  in the {P}oisson channel,'' {\em IEEE Trans.\ Inform.\ Theory}, 2012.

\bibitem{Verdu10IT}
S.~Verd\'u, ``Mismatched estimation and relative entropy,'' {\em IEEE Trans.\
  Inform.\ Theory}, vol.~56, pp.~3712--3720, Aug. 2010.

\bibitem{Weissm10IT}
T.~Weissman, ``The relationship between causal and noncausal mismatched
  estimation in continuous-time {AWGN} channels,'' {\em IEEE Trans.\ Inform.\
  Theory}, vol.~56, pp.~4256--4273, Sept. 2010.

\bibitem{TabPer12ISIT}
C.~G. Taborda and F.~P\'erez-Cruz, ``Mutual information and relative entropy
  over the binomial and negative binomial channels,'' in {\em Proc.\ IEEE Int.\
  Symp.\ Inform. Theory}, 2012.

\end{thebibliography}

\end{document}